\documentclass{article}

\usepackage{arxiv}
\usepackage{hyperref}       %
\usepackage{tabularx,booktabs} %

\usepackage{paralist}

\title{Qualification of Proof Assistants, Checkers, and Generators:
  Where Are We and What Next?\thanks{CC BY-NC-ND 4.0, the authors.
    This preprint was accepted for publication in the Science of
    Computer Programming journal and is now available under
    \url{https://doi.org/10.1016/j.scico.2023.102930}.}}

\author{Mario Gleirscher\\
  University of Bremen\\Bibliothekstrasse 5\\28359 Bremen, Germany\\
  \texttt{gleirsch@uni-bremen.de}\\
  \And
  Robert Sachtleben\\
  University of Bremen\\Bibliothekstrasse 5\\28359 Bremen, Germany\\
  \texttt{rob\_sac@uni-bremen.de}\\
  \And
  Jan Peleska\\
  Verified Systems International\\
  Am Fallturm 1\\
  28359 Bremen, Germany\\
  \texttt{peleska@uni-bremen.de}}

\makeatletter
\begin{document}
\maketitle

\begin{abstract}
  Cyber-physical systems, such as learning robots and other autonomous
  systems, employ high-integrity software in their safety-critical
  control.  This software is developed using a range of tools some of
  which need to be qualified for this purpose according to
  international standards.  In this article, we first evaluate the
  state of the art of tool qualification for proof assistants,
  checkers (e.g., model checkers), and generators (e.g., code
  generators, compilers) by means of a SWOT (Strengths, Weaknesses,
  Opportunities, Threats) analysis.  Our focus is on the qualification
  of tools in the three mentioned categories.  Our objective is to
  assess under which conditions these tools are already fit or could
  be made fit for use in the practical engineering and assurance of
  high-integrity control software.  In a second step, we derive a
  viewpoint and a corresponding range of suggestions for improved tool
  qualification from the results of our SWOT analysis.
\end{abstract}

\keywords{
  Deductive verification
  \and Model checking
  \and Certified compilers
  \and Cyber-physical systems
  \and Control software engineering
}

\section{Introduction}
\label{sec:introduction}

High-integrity software and control engineering is about the
engineering of electric, electronic, or programmable electronic %
systems required to operate at high degrees of dependability.
Dependability may include, for example, highly reliable components,
highly secure treatment of data and signals, and highly robust
enforcement of safe machine behaviour.  The more demanding such
requirements are, the more sophisticated methods, techniques, and
technologies are needed in control systems
engineering~\cite{Webster2014-GeneratingCertificationEvidence}.
Control software and systems subjected to such requirements
are used in many cyber-physical systems, such as intelligent collaborative
robots or autonomous vehicles.

The most demanding trustworthiness %
requirements can be tackled by the paradigm of \emph{formal design and
  verification}.  This paradigm fosters the use of formal logic
and mathematics in the construction and verification of key
engineering artefacts (e.g, hardware, software, models thereof).
Modern formal design and verification is highly automated, that is,
its results~(e.g., proofs, verdicts, models, code, executables) are
automatically
or interactively produced by tools and, thus, rely on the
correctness of these tools.
In this context, tools can be single points of failure with severe
impact on control system assurance and operation.  For example,
performance-increasing optimisations in untrustworthy compilers
\cite{bedinfranca:inria-00551370} have to remain switched off
until these compiler components are sufficiently verified.

Hence, the standards applicable to high-integrity systems development,
verification and validation (V\&V), and certification agree on the
fact that the trustworthiness of tools automating essential steps of
the life cycle process needs to be established in a way that is
comprehensible and can be checked by independent parties, for example,
the certification
authorities~\cite{iec61508,DO178C,CENELEC50128,iso26262-8}.  The
process for establishing this trustworthiness is called \emph{tool
  qualification~(TQ)}~\cite{DO178C} or \emph{tool
  validation}~\cite{CENELEC50128}.  We use the former term (or
`qualification' for short) to refer to the assurance of design and
verification tools or their results.  Qualification requirements in
certification~\cite{DO178C,DO330,iso26262-8,CENELEC50126a} are
justified by the fact that faulty tools automating parts of the
development process (e.g.,~compilers and other code generators) may
directly cause erroneous software to be deployed in the target system.
Faulty tools automating parts of the V\&V process (e.g.,~test
automation tools, model checkers, and proof assistants) may cause
errors in the target software to be overlooked.

\paragraph{Challenges.}

Reinforced by certification needs, the qualification of \emph{formal}
design and verification tools has gained traction in the formal
methods researcher and practitioner communities
\cite{DBLP:journals/dagstuhl-reports/CoferKSW15,%
  Wagner2017-FormalMethodsTool}, such as with proof
assistants~\cite{DBLP:conf/cade/ChihaniMR13}, model
checkers~\cite{Wagner2017-QualificationModelChecker},
compilers~\cite{AET2013-SCADESuiteKCG,kastner:hal-01399482}, abstract
interpreters, \textsc{Sat}isfiability/\-Modulo Theory (SAT/SMT)
solvers~\cite{DBLP:journals/jar/BlanchetteFLW18}, and conformance
testers~\cite{Brauer2012-EfficientTrustworthyTool,%
  Gleirscher2022-SoundDevelopmentSupervisors}.  Despite these
efforts, tool qualification is still not necessarily a key issue, even
in recent industry-focused
surveys~\cite{DBLP:journals/corr/abs-2003-06458}.  Where qualification
is seriously considered, such as in the aforementioned works, efforts
are either new~\cite{Wimmer2019-MuntaVerifiedModel}, limited in their
scope~\cite{Wagner2017-QualificationModelChecker}, or in an early
stage.  Most importantly, results are hence not yet broadly
transferred into high-integrity systems practice, and important
questions are still to be explored.

\paragraph{Research Hypothesis and Question.}

Motivated by the 
expectation
that formal design and verification will
become a key enabler on the pathway to trustworthy autonomous control, we 
focus on
the following question:
\begin{quote}
  What are the key issues in getting formal design and verification
  tools qualified, so that certification credit can be obtained for
  their generated results according to the industrial standards
  applicable
in  high-integrity control systems and software engineering?
\end{quote}
To limit the \emph{scope of our assessment}, we focus on proof
assistants, checkers, and generators as characterised below.  Hence, we
decided to exclude abstract interpretation and conformance testing tools
from our assessment, but take into account SAT/SMT solvers as far as
employed in the three focused categories.

\paragraph{The Three Tool Categories.} %

We hope that readers will find the following characterisation of formal
design and verification tools as a useful guide to the understanding
of our assessment below.
\begin{description}
\item[Proof Assistants] are formal verification tools that help
  engineers with the \emph{interactive} deductive proving of theorems,
  for example, about a system model or the correctness of control
  software.  Prominent examples for such tools include
  Coq~\cite{DBLP:series/txtcs/BertotC04},
  Isabelle~\cite{DBLP:books/sp/NipkowPW02},
  KeYmaera~X~\cite{Platzer2022-KeYmaeraXTutorial},
  Lean~\cite{Moura2015-LeanTheoremProver}, and
  PVS~\cite{Owre1992-PVSprototypeverification}.\footnote{Note that
    their code generation capabilities make these tools to be
    generators as well.}
\item[Checkers] are formal verification tools that allow one to
  \emph{automatically} check whether a $(\mathit{model},\mathit{property})$-pair
is element of a particular satisfaction relation ($\models$) or
whether a \emph{theorem} can be automatically deduced from a set of
\emph{axioms} given the \emph{inference rules} of a certain theory or
formal system.  Prominent examples for such tools include
\textsc{FDR4}~\cite{DBLP:journals/sttt/Gibson-Robinson16},
\textsc{Kind~2}~\cite{Champion2016-Kind2Model},
\textsc{nuXmv}~\cite{Cavada2014-nuXmvSymbolicModel},
\textsc{Prism}~\cite{Kwiatkowska2011-PRISM4.0Verification},
\textsc{ProB}~\cite{Leuschel2003-ProBModelChecker}, 
\textsc{Prolog}~\cite{DBLP:books/daglib/0083128},
\textsc{Spin}~\cite{Holzmann1997-ModelCheckerSPIN}, and
\textsc{Uppaal}~\cite{Behrmann2004-TutorialUppaal}.%
\footnote{As already mentioned and to limit the scope, we exclude
  solvers such as CVC4 and Z3.}
\item[Generators] are formal design tools 
  allowing to automatically translate or transform artefacts of an
  input language (e.g., C++, Mathworks
  Simulink\footnote{\url{https://www.mathworks.com/products/simulink.html}})
  into artefacts of an output language (e.g., an
  ARM executable,\footnote{\url{https://developer.arm.com/documentation/100166/0001/Programmers-Model}}
  SPARK~\cite{Barnes2003-HighIntegritySoftware}), while \emph{provably
    preserving} a range of correctness \emph{properties} (e.g., type
  and memory safety).  Prominent examples for such tools include
  certified compilers (e.g., CompCert~\cite{kastner:hal-01399482}) and
  model-to-code translators (e.g., the SCADE-to-X code
  generator~\cite{AET2013-SCADESuiteKCG}).
\end{description}

\paragraph{Overview.}

In Sec.~\ref{sec:tq-req}, we recapitulate key requirements on tool
qualification from the standards.  We assess the current situation by
means of a SWOT (Strengths, Weaknesses, Opportunities, Threats)
analysis in Sec.~\ref{sec:previous-work} and elaborate our viewpoint
with respect to the above question in Sec.~\ref{sec:what-needs-be}.  A
summary with concluding remarks is presented in Sec.~\ref{sec:conclusions}.

\section{What Do Standards Require?}
\label{sec:tq-req}

This section deals with the question:
\begin{quote}
  What do certification authorities (and industries) hope for in regard to
  tool qualification of proof assistants, checkers, and generators?
\end{quote}
There are several standards with guidance for tool
qualification that apply to the   three tool categories mentioned above.
According to our assessment, the guidelines in the avionic standard
RTCA DO-178C~\cite{DO178C} and its addendum RTCA DO-330~\cite{DO330}
currently specify the most comprehensive and strictest qualification
requirements (see Wagner et
al.~\cite{Wagner2017-QualificationModelChecker} for a summary).
The standard introduces
\emph{tool qualification levels} TQL-1 to TQL-5~\cite[Sec.~12.2.2]{DO178C}, \cite{DO330}.
A TQL corresponds to a \emph{set of qualification requirements} that
depends on (a)~the criticality of the target system to be developed, and (b)~the impact 
that tool failures could have on the software life cycle.  The impact is classified according to 
{\it Criteria~1}: the tool output becomes part of the target software, {\it Criteria~2}: the tool automates a verification activity, and its output is also used to eliminate or reduce parts of other   
verification activities and/or parts of the development, and {\it Criteria~3}: the tool automates a verification activity, so it can only overlook errors, but never inject errors into the target software.

Depending on the applicable TQL, the effort to be invested into the qualification of a tool varies considerably: the lowest level TQL-5 just requires \textbf{(R1)} configuration management for the tool software,  
\textbf{(R2)} documentation, \textbf{(R3)} verification, and \textbf{(R4)} validation of the tool operational requirements, and 
\textbf{(R5)} a test suite showing that these requirements have been adequately implemented and deployed in the 
production environment where the tool is used.  In
  contrast to that, the highest level TQL-1 requires \textbf{(R6)} a
fully documented software life cycle and \textbf{(R7)} verification
activities corresponding to those applicable to target system software
development for the highest design assurance level (DAL)~A~\cite{SAEARP-47541996,DO178C}, where software errors may lead to catastrophic consequences for the aircraft and its passengers~\cite[Annex~A]{DO330}.

With rigour and detail decreasing from TQL-1 to TQL-5, (R1) to (R5)
can include the following of certain more detailed development standards.
For example, (R5) may include integration tests to guarantee
the correct execution of the tool's object code in the run-time
environment.  (R6) can imply the documentation of tool requirements
and the tool architecture~(e.g.,~specifications at function or method
level).  Moreover, (R7) requires the verification of a tool's
behaviour against these requirements and the tool's robustness in
error situations, so that it is ensured that run-time errors are
detected.  The results from (R7) need to be checked for correctness
and completeness.  At the source code level, algorithmic correctness
might be shown by formally verifying the algorithms used by the tool.

Code generators producing C-code from control models, for example, need to be qualified according to TQL-1, if they generate DAL-A code, and no additional verification measures checking the consistency between code and model are applied. In contrast to this, model checkers and proof assistants automating a single step of the verification process are classified according to Criteria~3. Therefore, they need to be qualified only according to TQL-5, regardless of the criticality of the target system~\cite[Table~12.1]{DO178C}.

The situation is more subtle, if Criteria~2 apply. Suppose, for example, that 
a code verification tool has proven that no array boundary violations can occur in a software function.
If this verification
result %
is then used to conclude that the robustness tests concerning illegal array index values provided as function call parameters can be omitted, the tool would be classified according to Criteria~2, and this would result in TQL-4 if the target system code is associated with DAL-A or B.
The qualification effort needed for TQL-4 is considerably higher than that for TQL-5. It requires, in particular, an extensively documented   software life cycle which is hardly ever available for tools 
developed over decades in academic communities.

In summary, depending on the extent to which verification tools, such
as model checkers or proof assistants, \emph{replace manual
  verification or testing}, they can either be subjected to just TQL-5 or,
for example, for airborne DAL-A software, even to TQL-4 if they automate
critical parts of the V\&V process.

The EN~50128 standard for railway applications~\cite{CENELEC50128}
mandates a scheme similar to DO-178C, also with three tool categories,
T1, T2, and T3.  The integrity level of the target system is, however,
left implicit.  According to the supplement
EN~50126~\cite{CENELEC50126a}, T3 with most rigorous implications on
qualification applies to tools, such as code generators and compilers,
that can insert errors into train system components.  T2 and T1
represent less rigorous requirements (akin to TQL-4 and~5) to be
followed, for example, by tools unable to insert faults but prone to
overlook faults (e.g., verification tools) and tools neither
automatically inserting nor detecting faults (e.g., editing tools).

It should be emphasised that tool qualification is not a once-and-for-all activity~\cite[Sec.~12.2.1]{DO178C}: the standards stress that the qualification needs to be re-assessed with every new product development undertaking, because different target system characteristics may require different tool properties. The product-specific tool qualification, however, can usually rely on reusable tool components applicable to every target system type, so that \emph{tool qualification kits} can be prepared as templates facilitating the target system-specific qualification. 

Tool qualification according to these requirements
cannot guarantee complete correctness of a tool.  This, however, is
not the intention of the standards defining them. Instead, following
the specified tool qualification process guarantees
\emph{accountability} and \emph{liability}: In case of an undetected
tool malfunction leading to a severe failure in the target system, the
artefacts generated during tool qualification allow one to determine
whether the tool has not been adequately verified. If this is the
case, the system developers having applied and qualified the tool are
liable for the consequences of the target system failure. Conversely,
if tool qualification has been comprehensively performed according to
the requirements of the applicable standard, the developers avoid
liability by arguing that all measures that can be ``reasonably
expected'' to ensure tool correctness have been performed.

\section{What Has Been Done?} 
\label{sec:previous-work}

Following the SWOT analysis
guidelines~\cite{Piercy1989-MakingSWOTAnalysis}, we present some
\emph{Strengths}~(Sec.~\ref{sec:strengths}), including achievements in
the qualification of proof assistants, checkers, and generators.
Likewise, we summarise \emph{Weaknesses}~(Sec.~\ref{sec:weaknesses})
known from the literature and from our academic and industrial
experience.  Finally, we indicate
\emph{Opportunities}~(Sec.~\ref{sec:opportunities}) and
\emph{Threats}~(Sec.~\ref{sec:threats}) suggesting new research areas
and possibilities for enhancing existing work.  Below, we identify
each SWOT item with its initial---\textbf{S}, \textbf{W}, \textbf{O},
or \textbf{T}---and a sequence number.

\subsection{Strengths} %
\label{sec:strengths}

One can observe that the tools in the considered categories (e.g., checking,
deduction, compilation) have become quite powerful and some of these
tools are increasingly easy to use, not only from the viewpoints of
their developers but also suggested by expert users in academia
\cite{Garavel2020-2020ExpertSurvey} and industry
\cite{Gleirscher2020-FormalMethodsDependable}.
Hence, this section addresses the question:
\begin{quote}
  What are the key achievements in the qualification of proof
  assistants, checkers, and generators?
\end{quote}
Regarding the qualification of proof assistants, 
\begin{inparaenum}
\item[\textbf{(S1)}] proof terms or objects, capturing basic inferences constituting
  a mechanised proof, can be
  exported~\cite{DBLP:conf/tphol/BerghoferN00,wenzel2021isabelle} to
  enable the checking of proofs performed in them by additional proof
  checkers~\cite{DBLP:journals/fmsd/Wong99,DBLP:conf/cade/NipkowR21,%
    DBLP:journals/pacmpl/SozeauBFTW20,DBLP:journals/jlap/Abrahamsson20,%
    anssi_inria2021coq}.
\item[\textbf{(S2)}] Some assistants even allow one to reason about their
  underlying logics or implementations,
  enabling self-formalisation efforts in order to raise confidence in the their correctness~\cite{DBLP:conf/cade/Harrison06,DBLP:journals/jar/KumarAMO16,DBLP:conf/cade/NipkowR21}.
  Such efforts have led, for example, to the HOL~Light proof assistant, mechanically
  verified down to the level of machine
  code~\cite{DBLP:conf/itp/AbrahamssonMKS22,DBLP:conf/itp/Myreen21}.
\end{inparaenum}

In the context of the qualification of checkers,
\begin{inparaenum}
\item[\textbf{(S3)}] 
  Bendisposto et al.~\cite{Bendisposto2014-Whowatcheswatchers} report
  on the effort of validating the ProB model checker as an EN~50128
  class T2 tool.  Bendisposto et al.'s qualification argument includes
  a proven-in-use claim, extensive coverage-oriented testing, static
  analysis, and cross-checking with another B parser (Atelier~B
  bcomp~\cite{ClearSy2009-AtelierBUser}).
  In contrast, Wagner et
  al.~\cite{Wagner2017-QualificationModelChecker} argue that it is
  sufficient for a model checker (e.g.,
  \textsc{Kind}~2~\cite{Champion2016-Kind2Model}) to achieve
  TQL-5 (similar to T1).  After deriving 111 requirements for that
  checker and further requirements for a proof checker, they perform
  corresponding (language coverage) tests and peer reviews according
  to DO-330~\cite{DO330} (cf.~Sec.~\ref{sec:tq-req}).

  Wagner et al.\ demonstrate the 
\item[\textbf{(S4)}] %
  independent verification of \textsc{Kind}~2's
  output with a simple-to-qualify proof checker (i.e., Check-It based
  on the LFSC\footnote{Logical Framework with Side
    Conditions}~\cite{Stump2012-SMTproofchecking}) based on proof
  certificates of an SMT solver.  In their study, the CVC4
  solver~\cite{Barrett2011-CVC4} provides a $k$-induction-based
  certificate that unsafe states cannot be
  reached~\cite{Mebsout2016-ProofcertificatesSMT}.  Wagner et al.\
  then apply their TQL-5 argument to Check-It instead.
  For reachability in timed automata, Wimmer and {von
    Mutius}~\cite{Wimmer2020-VerifiedCertificationReachability} show
  how proof certificates can be represented as compressed sets of
  abstracted symbolic states.
  In addition to certificates for the class of safety
  properties~\cite{Mebsout2016-ProofcertificatesSMT,
    Wimmer2020-VerifiedCertificationReachability},
  liveness certificates can also be generated in SAT-based linear
  temporal logic (LTL) model
  checkers~\cite{Griggio2021-CertifyingproofsSAT} and for timed
  automata model checkers~\cite{Wimmer2020-CertifyingEmptinessTimed}.
  
  Furthermore, 
\item[\textbf{(S5)}] %
  some core algorithms of a variety of model checkers have
  been verified using proof assistants, for example, \textsc{Prism}'s
  probabilistic computation tree logic checking algorithm \cite{Hoelzl2012-VerifyingpCTLModel}, a
  \textsc{Uppaal}-compatible formalisation of timed automata
  checking~\cite{Wimmer2018-VerifiedModelChecking} together with a
  multi-tier qualification argument
  \cite[Sec.~4.2]{Wimmer2019-MuntaVerifiedModel}, and a proof of the
  critical LTL-to-B\"uchi automaton
  translation~\cite{Schimpf2009-ConstructionBuchiAutomata}, the three
  approaches using Isabelle/HOL.
\end{inparaenum}

Concerning the qualification of generators,
\begin{inparaenum}
\item[\textbf{(S6)}] the optimising, semantics-preserving C compiler
  \textsl{CompCert}, formally verified in Coq, has been successfully
  qualified for use in the development and certification according to
  IEC~60880 of a high-integrity control system in the nuclear power
  domain \cite{kastner:hal-01643290}.

  Underlining DO-178C's testing-driven qualification,
\item[\textbf{(S7)}] Taft et
  al.~\cite[Sec.~5]{Taft2017-BuildingTrustModel} suggest that TQL-1
  for a Simulink-to-SPARK or -MISRA~C generator can be achieved by
  integrated unit testing with sufficient coverage of the generator
  input language.  Here, the language of Simulink models is covered by
  traversing its block grammar rules.  The authors describe how
  input-language coverage can drive test requirements (i.e., input
  equivalence classes for code units of a generator), the creation of
  assertion checkers with oracles (i.e., outputs of the units expected
  for certain inputs to these units), and, ultimately, test coverage
  (i.e., sufficiently covering the input/output behaviour of each of
  the units).
  
\item[\textbf{(S8)}] The commercial ANSYS SCADE Suite creates code
  from SCADE models and offers a TQL-1-conforming tool qualification
  kit~\cite{DBLP:conf/oopsla/Colaco20,AET2013-SCADESuiteKCG}.
  Moreover, the generated object code can be verified against the
  generated C-code.  Both tool qualification and object code
  validation rely on tests achieving Modified Condition/Decision
  Coverage~(MC/DC).  These suffice according to DO-178B to obtain
  TQL-1 for the tool and to verify the production object code for
  DAL-A.
  In other cases, 
\item[\textbf{(S9)}] code generation can be certified (in class T3)
  when using diverse implementations with code produced by at least two
  independent code generators and the resulting executables running on
  diverse hardware~\cite{Leuschel2022-PrivateCommunication}.
\end{inparaenum}

In summary, there are several efforts to reduce the code base
responsible for proof construction to a minimal \emph{trusted core} of
verifiable routines.  Examples include the concentration of
\textsc{Kind}~2's qualification on the verification of
CheckIt~\cite{Wagner2017-QualificationModelChecker} as well as the
focus of qualification efforts in Isabelle on its kernel, which is
orientated towards the compact Logic-of-Computable-Functions~(LCF)
principle.
On the side of code generators, Yuan et
al.~\cite{Yuan2022-EndEndMechanized} follow a refinement-based
approach (proving a forward simulation relation) to certify
semantics-preserving C-code generation from a functional program
previously security-verified in Coq.  They establish confidence in
their certification by a careful minimisation and validation of the
trusted computing base.
On the other side, the approaches to qualifying proof assistants and
model checkers, desired to be more
trustworthy~\cite{Necula1998-designimplementationcertifying}, seem to
focus on artefacts (e.g., validating the tool's result by an
additional checker) rather than the tools themselves (e.g., testing
the tool and verifying its algorithms).

\subsection{Weaknesses} %
\label{sec:weaknesses}

This section is centred around the question:
\begin{quote}
  Why can tool qualification sometimes fail?  What can go wrong when
  using a proof assistant, checker, or generator in an industrial
  certification setting?
\end{quote}
\begin{inparaenum}
\item[\textbf{(W1)}] Arguments based on the soundness of the
underlying logic of a proof assistant may be invalidated if
user-defined axioms are
added~\cite[Sec.~3]{DBLP:journals/jfrea/Adams16}. Some
assistants allow suspension of certain proof obligations in
their interactive modes~\cite{wenzel2021isabelle} or by disabling
guard conditions~\cite{DBLP:conf/esop/MonniauxB22}.
\item[\textbf{(W2)}] Moreover, proof checkers developed to check proofs of the same proof
assistant they have been implemented and verified in, such
as~\cite{DBLP:conf/cade/NipkowR21,DBLP:journals/pacmpl/SozeauBFTW20},
may not be sufficiently diverse to be applied in the independent
checking of proofs.
\item[\textbf{(W3)}] \label{wea:fallback-to-manual} Importantly, if proof assistants
are not qualified and their proofs are not checked by a sufficiently
diverse\footnote{Some regulators using older versions of IEC~61508
  were not even positive about the mitigation of a failure in a
  proof assistant (e.g., PVS) via the use of a different
  tool~\cite[Sec.~2]{DBLP:conf/vstte/Lawford16}.}  proof
checker---that is, if they constitute a single point of
failure---then it may be necessary to fall back to a manual
(independent) repetition of these
proofs~\cite[Sec.~2]{DBLP:conf/vstte/Lawford16}.
\end{inparaenum}

\begin{inparaenum}
\item[\textbf{(W4)}] For formal method applications,
  Taft et al.~\cite{Taft2017-BuildingTrustModel} state that ``the
  co-development of formal specifications, proofs, and program code
  raises significant `common mode' concerns''.
\item[\textbf{(W5)}] For model checkers, Wagner et
  al.~\cite{Wagner2017-QualificationModelChecker} stress that common
  cause errors in an unqualified model checker and a qualified proof
  checker are easy to be overseen.
\end{inparaenum}

Concerning the qualification of generators, 
\begin{inparaenum}
\item[\textbf{(W6)}] code generated from provably correct types (i.e.,
  specifications of data structures and algorithms) may not be ensured
  to conform to those types if the generation mechanism is
  unqualified.  This can be partially tackled by verified code
  generators such as~\cite{belanger2019certified}
  and~\cite{DBLP:conf/esop/HupelN18} for Coq and Isabelle/HOL.
\item[\textbf{(W7)}] Though conforming to the qualification
  requirements discussed in Sec.~\ref{sec:tq-req}, the form of
  generator testing suggested in
  \cite[Sec.~5]{Taft2017-BuildingTrustModel}, even if applied with
  MC/DC tests to achieve TQL-1 (see also S8), is not a valid
  replacement for rigorous proofs of tool correctness or equivalence
  checks of generated code against models.  
\end{inparaenum}

Furthermore, we need to acknowledge that qualification is supposed to
not only provide evidence for the correctness of the tools in
isolation but also to show the suitability of these tools in regard to
the requirements of the particular development and V\&V projects.
Appreciating the state of the art, however, it needs to be said, that
\textbf{(W8)} the mentioned tools have largely been used in
environments where certification is not required (e.g., in academic
settings).  Moreover, certification authorities have still rarely
considered formal design and verification as key activities and their
results as key artefacts in industrial assurance cases.  It might thus
be unsurprising that tool
providers (e.g., research teams) have only given limited attention to
a comprehensive industry-strength qualification.

\label{sec:transl-input-valid}
It is important to note that tool qualification is not
about assuring that the input to a proof assistant (i.e., a type and
a theorem) or model checker (i.e., a model and a property) is actually
valid.  Nevertheless, the \textbf{(W9)} crucial issues of
\emph{translation} and \emph{input validation} can have a tremendous impact
on tool qualification.

\label{sec:translation}
As observed by Cofer~\cite[p.~24]{Cofer2015-Youkeepusing}, the risks
stemming from a \emph{translation} required to use a proof assistant
or model checker can in certain cases outweigh that tool's benefits,
in comparison with the often less involved default translations.

Default translations reflect what might occur in any safety-critical
development and usually encompass, for example, the necessary
transformation of requirements into an implementation, the derivation
of a test suite from these requirements, the occasional translation of
critical requirement fragments into useful but often free-style
mathematical sketches.
The \emph{additional translation} required to use a proof assistant or
model checker usually includes (i)~a careful transcription and
abstraction of the requirements and the domain description into a
$(\mathit{model},\mathit{property})$-pair,\footnote{For the sake of
  simplicity, we consider a $(\mathit{model},\mathit{property})$-pair
  for a model checker to be the dual of a
  $(\mathit{type},\mathit{theorem})$-pair for a proof assistant.} and
then (ii)~an alignment of the verified model with, or a translation
into, an implementation.

The aforementioned risks may thus include, for example, issues in
understanding an additional model in another language or the costs of
aligning that model with the existing descriptions by tracing backward
through the corresponding translations and abstractions.

\label{sec:input-validation}
Concerning \emph{input validation}, model checkers and proof
assistants do not usually help one in figuring out how much a checked
property or deduced theorem contributes to a proof
obligation~\cite[Sec.~3.1.2]{DBLP:journals/jfrea/Adams16}.  Overall,
translation- and input-related issues threaten the essential
correctness assertions that (i) the
$(\mathit{model}, \mathit{property})$-pair reflects the domain and the
intended requirements, and (ii) the specified property is strong
enough to rule out undesirable models.

\subsection{Opportunities}
\label{sec:opportunities}

Based on previous studies
\cite{Gleirscher2020-FormalMethodsDependable,%
  Garavel2020-2020ExpertSurvey}, we are confident that formal design
and verification tools have good potential to improve the best
practices in the development and assurance of control software.  Here,
we focus on the question:
\begin{quote}
  Building on the strengths (Sec.~\ref{sec:strengths}), which
  opportunities are there for getting these tools qualified?
\end{quote}
\begin{inparaenum}
\item[\textbf{(O1)}] A qualification of the Isabelle proof assistant
  would lead to a significant number of software products verified by
  means of Isabelle to become certifiable. To provide an example,
  the seL4 case study applies Isabelle comprehensively to
  verify an operating system kernel~\cite{DBLP:conf/sosp/KleinEHACDEEKNSTW09}, focusing in particular
  on security properties such as the enforcement of strong isolation
  between applications run on top of it, intending it to serve as the
  foundation of trustworthy systems
  \cite{Klein2018-Formallyverifiedsoftware}.  seL4 is not yet
  certified~\cite{VanderLeest2018}: As suggested
    in~\cite[Talk 3.1]{DBLP:journals/dagstuhl-reports/CoferKSW15},
  tool qualification of Isabelle is the essential prerequisite to
  obtain certification credit for the correctness proofs developed for
  the kernel.  Thus, a \emph{reference qualification} for Isabelle,
  yet missing, could increase the motivation to qualify other proof
  assistants as well.
  
\item[\textbf{(O2)}] Regarding the design of proof assistants, the
  simplicity of the LCF approach underlying many modern proof
  assistants could be further exploited
  \cite{DBLP:journals/jfrea/Adams16,%
    DBLP:journals/corr/abs-2003-06458}.  This approach prescribes an
  inference kernel as the sole soundness-critical component,
  performing a limited number of primitive inference rules and
  operations to expand theories.  From studying this kernel, we could
  learn how to qualify the cores of proof assistants by sufficient
  unit testing and turn them into trusted components.
  
\item[\textbf{(O3)}] \label{opp:fallback-to-manual} Accepting
  potential fallbacks to manual proofs
  (W3) %
  as highlighted
  in~\cite[Sec.~2]{DBLP:conf/vstte/Lawford16}, tool-informed manual
  proofs could still be an improvement over entirely manually
  performed proofs.  In low-criticality contexts, we could then think
  about using proof assistants for that purpose even without
  qualification.
\item[\textbf{(O4)}] \label{opp:proof-audit}%
  Moreover, the qualification of proof assistants could benefit from
  extended guidance on how to use them to address the requirements of
  specific standards.  For example, Bertot et al.~\cite[Sec.~3.3]{bertot2020coq,anssi_inria2021coq}
  provide recommendations on how to employ Coq in the context of the Common
  Criteria~(CC) standard (e.g., a restricted configuration of Coq when used
  for the demanding CC~evaluation assurance levels EAL 6+/7).
\end{inparaenum}

Concerning the qualification of checkers and in the context of TQL-5
qualification, \textbf{(O5)} Wagner et
al.~\cite{Wagner2017-QualificationModelChecker} spread optimism by
stating that ``the guidance of DO-330 does not require any activities
that are especially difficult or costly for qualification of a model
checker.''  So, there is an opportunity for other model checkers to
follow their approach.

In regard to the qualification of generators, 
\begin{inparaenum}
\item[\textbf{(O6)}] certified
  compilers, such as CompCert, can confidently raise the level of abstraction of proof
  efforts, given the supported language fragment fits the practical
  needs.  
  For instance, if it can be certified that a compiler preserves the semantics of a given program between source code and compiled code, then it becomes possible to analyse the source code and soundly transfer the results to the compiled code. This reduces the need for additional analysis of the compiled code and costly reviews on whether the compiled code corresponds to the source code \cite{brahmi:hal-01708332}.  
  A nearby opportunity would be to try to achieve TQL-1 of the trusted
  computing base of code generators using semantically-integrated
  refinement proofs as, for example, shown
  in~\cite{Yuan2022-EndEndMechanized}.   An alternative solution to
  dealing with simpler generators could be to use a complete testing
  approach~\cite{Gleirscher2022-SoundDevelopmentSupervisors}.
  
\item[\textbf{(O7)}] %
  Here, the code generation capabilities of proof assistants constitute a particularly motivating target for tool qualification. 
  If they could be qualified to generate code from specifications developed in proof assistants so that the generated code retains properties established over the specifications via the proof assistant, this would foster the qualification of a large number of already existing tools developed via proof assistants, including LTL model
  checkers~\cite{DBLP:conf/cav/EsparzaLNNSS13}, SAT
  solvers~\cite{DBLP:journals/jar/BlanchetteFLW18}, and 
  compilers~\cite{kastner:hal-01643290,DBLP:conf/esop/HupelN18}.  
  For instance, Bendisposto et
  al.'s~\cite{Bendisposto2014-Whowatcheswatchers} aim is to achieve a
  validation of the ProB generation facilities as an EN~50128 class T3
  tool.
\end{inparaenum}

\label{opp:diverse-tool-use} Following the idea of
artefact-based qualification, \textbf{(O8)} we could improve the
support of redoing interactive proofs or automatic checks in diverse
proof
assistants~\cite{DBLP:conf/nfm/Hurd11,DBLP:journals/jsc/GauthierK19,DBLP:conf/itp/KellerW10}
and model checkers.

We could further \textbf{(O9)} mitigate the risk of a poor
cost-benefit ratio from using formal design and verification tools.
Regarding the required \emph{additional translations}~(W9), two
types of benefits are important to be considered:
\begin{inparaitem}[]
\item a short-term benefit from early fault detection and
\item a long-term benefit from having reusable domain models
  and theories.
\end{inparaitem}
These benefits could be strengthened by methodologies aiding in the
translation between requirements or assurance documents and the input
languages of model checkers and proof assistants.  See the two recent
examples in
\cite{Farrell2022-FRETtingRequirementsFormalised,Foster2020-IntegrationFormalProof}.

Part of the \emph{input validation} problem has been addressed by
\begin{inparaitem}[]
\item \emph{coverage metrics and sanity
    checks}~\cite{Chockler2006-CoverageMetricsFormal}, for example,
  the automatic identification of easily breakable preconditions
  leading to vacuously true implications,
\item \emph{warnings about problematic proof directives} (cf.~W1), as
  reported, for example, to Isabelle users~\cite{wenzel2021isabelle},
  and
\item \emph{model-in-the-loop testing}, for example, to make sure that
  the model captures domain phenomena in such a way that useful
  properties can be specified.
\end{inparaitem}
Along with these measures, tool-specific debugging techniques
can help us to increase the confidence and trust in formal design and
verification tools.
Furthermore, informal-to-formal translations and input validation
could be controlled with frameworks such as
Isabelle/DOF~\cite{Brucker2019-IsabelleDOFDesignImplementation},
demonstrating direct support of requirements traceability within proof
environments.

Finally, \textbf{(O10)} we expect that proof assistants, checkers, and
generators will increasingly be used in high-integrity system
assurance, so that their qualification will gain increasing attention.

\subsection{Threats} %
\label{sec:threats}

In this section, we focus on the following question:
\begin{quote}
  Which obstacles to the qualification of formal design and
  verification tools are to be expected?
\end{quote}
Weaknesses refer to problems observed
or known with the existing state of the art.  Threats focus on
challenges to cope with and risks to reoccur or to be faced, in
particular, when trying to pursue the opportunities listed in
Sec.~\ref{sec:opportunities}.

Concerning the qualification of proof assistants, two threats
  may limit the benefits of tool qualification guidelines~(O4):
\begin{inparaenum}
\item[\textbf{(T1)}] First, the code bases of even the cores of Isabelle and Coq
exceed 10,000 lines of code, and future versions of both tools could
exhibit soundness flaws, as detailed for earlier versions
in~\cite[Sec. 3.2]{DBLP:journals/jfrea/Adams16}. 
\item[\textbf{(T2)}] Moreover, components outside of these cores may still need to be
qualified to ensure that a proof assistant checks the proofs the user
expects it to check.  This qualification, in particular, includes printing and
parsing capabilities, where many assistants allow
\begin{inparaenum}[(i)]
\item syntactic manipulations that could mislead users about what
  theorems they are actually
  proving~\cite{DBLP:journals/entcs/Wiedijk12} or
\item to print theorems in ways that, when parsed again, result in
  different theorems.
\end{inparaenum}
This aspect has been addressed explicitly only by a few tools, such as
HOL Zero~\cite{DBLP:conf/itp/Adams16}, which has been developed for
this very purpose.

\item[\textbf{(T3)}] \label{wea:proven-in-use} The well-structured qualification case
in \cite[Sec.~4.2]{Wimmer2019-MuntaVerifiedModel} for a timed
automaton checker code base contains the clause ``it is widely
accepted within the community that Isabelle/HOL only admits valid
theorems (at least on the user level).''  As a \emph{proven-in-use}
argument, even a well-justified observation like this needs to be
underpinned by detailed evidence   for a certification authority, as indicated
by Adams~\cite[Sec.~3.1.2]{DBLP:journals/jfrea/Adams16}:
According to the applicable standards, a proven-in-use argument requires    a comprehensive documentation of its \emph{service history}~\cite[12.3.4]{DO178C}. This documentation comprises a complete history of configuration management and product failure tracking.  Such a history will not usually be available for verification tools developed in the academic communities.  Moreover, obtaining certification credit for such a tool in a particular V\&V campaign requires that the tool has been used extensively in the past and in domains that are similar to the actual system to be verified.  Providing a history of successful use will hardly be feasible for tools developed in academia, since the tool applications---though extensive---are usually on examples stemming from open-source developments or significant academic challenges that were not directly based on V\&V campaigns for safety-critical industrial systems. 
\end{inparaenum}

Regarding the qualification of checkers, in particular when
  pursuing TQL-5~(O5) or proof replay~(O8),
\begin{inparaenum}
\item[\textbf{(T4)}] it can be challenging to explain the indirect approach to
  assurance with additional proof checkers to certification
  authorities~\cite{Wagner2017-QualificationModelChecker}.
\item[\textbf{(T5)}] The effort of qualifying a proof checker for
  TQL-5 can, although not difficult (W5), be
  nearly as high as the effort for qualifying the model checker itself
  for TQL-5~\cite{Wagner2017-QualificationModelChecker}.  Similar
  efforts could raise two questions: Depending on the capabilities of
  the proof checker, could the latter be used as the model checker in
  the first place?
  And, relating to~(O3), if a tool is only used to double-check manual
  proofs, do we need tool qualification at all?
  On the one hand, Wagner et
  al.~\cite[Sec.~6]{Wagner2017-QualificationModelChecker} discuss how
  TQL-4 might require a much higher effort than TQL-5.  On the other
  hand, it remains to be seen how often TQL-4 applies to tools for
  verifying DAL-A code if these tools do not replace large parts of
  the DAL-A verification procedure.
\item[\textbf{(T6)}] In addition, the format and size of proof objects and
  certificates for large systems could make their validation
  difficult.
\end{inparaenum}

Regarding the qualification of generators, \textbf{(T7)} as indicated
by Yang et al.~\cite{Yang2011-Findingunderstandingbugs}, even
compilers having undergone serious certification
efforts~\cite{kastner:hal-01399482} might still have critical flaws,
threatening the preservation of the semantics~(O6).  Hence,
it is crucial that a tool qualification case makes explicit the exact
circumstances and the code for which certification credit was
obtained.

The following threats pertain to any kind of tool.
\begin{inparaenum}
\item[\textbf{(T8)}] As an organisational aspect and human
  factor, there could be a lack of trust on the side of certification
  assessors into V\&V approaches based on formal methods and
  supporting tools.  This lack of trust could hinder qualification.
\item[\textbf{(T9)}] Moreover, tools from the three categories discussed in this
  article face the usual risk of their code bases getting too complex
  to be qualified for a targeted TQL.  Similarly, discontinuities in
  tool maintenance (e.g., missing or new developers) could hinder
  re-qualification.  Additionally, changes of both, tools or standards
  (e.g., DO-178C), could trigger costly re-qualification.
\end{inparaenum}  

\section{Our Viewpoint: What Could Be Done Next?} %
\label{sec:what-needs-be}

This section is an attempt to suggest answers to the following three
questions:
\begin{quote}
  How can we handle the weaknesses (Sec.~\ref{sec:weaknesses})?  How
  can we use the opportunities (Sec.~\ref{sec:opportunities}) to the
  maximum benefit?  What could we do to mitigate the threats
  (Sec.~\ref{sec:threats})?
\end{quote}
Our \emph{viewpoint} is that a repetitive and exhaustive formal
verification of tools (e.g., Isabelle beyond its core or the
collection of \textsc{Prism}'s algorithms) is likely to remain
difficult if not infeasible, both economically and organisationally.
Some tools have just grown too complex and might naturally remain
black-boxes, even to their expert users.

In the context of artefact-based tool qualification, our SWOT analysis
in Sec.~\ref{sec:previous-work} and our experiences with the state of
the art suggest several tasks to be considered for a successful
qualification of formal design and verification tools.
  
\begin{inparaenum}
\item[\textbf{(V1)}] %
  Based on the widely accepted \emph{de Bruijn} criterion for
  checkers~\cite{Barendregt2005-challengecomputermathematics}\footnote{``A
    mathematical assistant satisfying the possibility of independent
    checking by a small program is said to satisfy the de Bruijn
    criterion.''  \cite{Barendregt2005-challengecomputermathematics}}
  and on Pnueli's idea of generator output
  validation~\cite{DBLP:journals/sttt/PnueliSS98,DBLP:conf/birthday/PnueliSS99},
  we  
  advocate
  the enhancement or replacement of tool
  verification by artefact-based tool qualification.
  Successfully qualified artefacts could replace the qualification of
  the tools producing these artefacts.  The conditions stated in
  DO-178C~\cite[Sec.~12.2]{DO178C} will then be imposed on the
  artefact validation tools, ranging from utilities for computing
  ``checksums'' (and other properties indicative of the validity of
  the output) to preferably simple proof checkers or proof certificate
  analysers certifying   the validity of the output.
  This form of qualification is not only increasingly supported by the
  research community, as discussed in Sec.~\ref{sec:previous-work}, but
  would also address the principle of \emph{Explainability} in the
  Manifesto for Applicable Formal
  Methods~\cite{Gleirscher2021-ManifestoApplicableFormal}.

\item[\textbf{(V2)}] The comprehensive testing according to DO-178B/C for the
  qualification of a complete model checker or proof assistant might
  also be infeasible.  As discussed in Sec.~\ref{sec:threats}, 
  a proven-in-use argument will not be acceptable from the standards' perspective.
  \label{sug:diverse-tool-use} Hence, following the suggestion in
  \cite[Sec.~12.3.2.4/5]{DO178C}, we again stress the achievement of
  \emph{redundancy} of proof results through the use of \emph{diverse
    tools}, as also suggested by Fantechi and
  Gnesi~\cite{Fantechi2011-AdoptionModelChecking} and successfully applied, for example, by
  Parillaud et al.~\cite{DBLP:conf/rssrail/ParillaudFB19}.
  
\item[\textbf{(V3)}] \label{sug:std-proof-cert} More specifically,
  trust in tools could be increased by improving proof certificates.
  In particular, the application of diverse tools to the same problems
  could be facilitated by developing and adopting shared formats
  analogous to the TPTP\footnote{Thousands of Problems for Theorem
    Provers} family of exchange
  formats~\cite{DBLP:journals/jar/Sutcliffe17}.  At the conceptual
  level, Chihani et al.~\cite{DBLP:conf/cade/ChihaniMR13} characterise
  core elements of proof certificates to be defined and shared among
  provers and solvers.  For model checkers, Beyer et
  al.~\cite{Beyer2016-CorrectnessWitnessesExchanging} explore a
  graph-based shared format for correctness witnesses\footnote{In
    model checking, \emph{counterexamples} witness the violation of a
    specification (e.g., a temporal logic formula) and
    \emph{witnesses} are models of a specification.  In contrast,
    \emph{correctness witnesses} compactly represent complete sets of
    witnesses, that is, invariants certifying safety of a model.}
  produced by these tools.

\item[\textbf{(V4)}] We think that, for further industries to gain
  trust in the mentioned tools, early \emph{qualification cases}
  should be open-source rather than
  undisclosed
  and only shown to certification
  authorities (e.g., the undisclosed case of the
  DO-178C/TQL-1-certified SCADE code
  generator~\cite{AET2013-SCADESuiteKCG}).
  Generally, we should build up lasting trust in new technologies of
  any kind by making reference cases open-source (e.g., the seL4
  proofs~\cite{Klein2018-Formallyverifiedsoftware,VanderLeest2018})
\end{inparaenum}

\begin{inparaenum}
\item[\textbf{(V5)}] \label{sug:proof-audit} Specific guidance for the \emph{auditing
    of mechanised proofs} could be developed in order to simplify and
  regulate this process.  This could include both tool-specific
  guidance such as the need to check for the suspension of proof
  obligations or disabling of guards, which may affect soundness, as
  well as general guidelines for the examination of theorems and their
  dependencies as suggested
  in~\cite[Sec.~3]{DBLP:journals/jfrea/Adams16}.  Analogously,
  guidelines for the use of proof assistants such
  as~\cite{anssi_inria2021coq} %
  for Coq could be extended to other tools and checked against in
  proof audits.  Considering the possibility of tool-informed manual
  proofs, proof auditing could thus foster the use of mechanised
  proofs in qualification efforts instead of only relying on proof
  assistants themselves.  Hence, we would like to advocate
  tool-informed over fully manual proofs.  Human-friendly proof
  languages such as Isabelle/Isar~\cite{wenzel2021isabelle} can enable
  such a form of interaction.\footnote{In a recent comment on
    improving the interoperability between proof assistants, Paulson
    stresses the role of a human-friendly proof language in the
    porting of proofs at a conceptual level rather than at the low
    level of mechanical inferences.  See
    \url{https://lawrencecpaulson.github.io/2022/09/14/Libraries.html}.}
  Generally, an improved tool support for input validation of and
  translation between artefacts~(W9, O9) would devitalise the
  occasional ``garbage-in/garbage-out'' argument against formal design
  and verification tools.

\item[\textbf{(V6)}] \label{sug:truly-indep-proof-chk} To prevent us
  from constructing circular qualification
  arguments,\footnote{Verifying a proof checker in the very proof
    assistant whose proofs it is to check may raise issues about
    circular verification.} an alternative would be to strive for
  sufficiently diverse and independent proof checkers, for example,
  one for Isabelle/HOL verified and implemented in Coq and vice versa.
\end{inparaenum}

\textbf{(V7)} Given a verified model checker (e.g., as shown in
\cite{Wimmer2018-VerifiedModelChecking,
  Hoelzl2012-VerifyingpCTLModel}) or a simple proof checker, and apart
from the issues of an additional translation (as described in
Sec.~\ref{sec:translation} and to be addressed by, e.g., model-in-the-loop
simulation), we think that there are no hard-to-overcome obstacles to
getting a model checker qualified up to TQL-5.

Moreover, \textbf{(V8)} proof certificates could complement certified
compilers with certifying
compilers~\cite{Necula1998-designimplementationcertifying}.

In a survey of benefits and limits of tool qualification according to
DO-330 and DO-178, Ibrahim and
Durak~\cite{Ibrahim2021-StateArtSoftware} allude to the trend ``of not
qualifying development tools [but rather] qualifying verification
tools''.  To avoid costly re-qualification (T9), they suggest
\textbf{(V9)} the definition of use cases within which a tool can be
transferred from one project to another.  This idea could perhaps be
adopted in the qualification of formal verification and design tools.

\section{Summary}
\label{sec:conclusions}

In this work, we have assessed some key issues for getting proof
assistants, checkers, and generators qualified according to the
applicable standards.  Table~\ref{tab:swot:tq-fm} provides a
  summary of our findings from the SWOT analysis as well as the
suggestions accompanied with our viewpoint.
We hope that our discussion contributes to the next generation of
qualified formal design and verification tools in order for them to be
used in the development and certification of control software used in
cyber-physical systems, such as intelligent robots and autonomous
systems~\cite{Webster2014-GeneratingCertificationEvidence}.
Finally, a wider assessment of the field under discussion could
include (i) tools for abstract interpretation, further solvers and
checkers, and (ii) proof-carrying code could be similarly discussed as
a form of proof certificates.

\begin{table}[t]
  \caption{Summary of our SWOT analysis of tool qualification in
    formal methods applications
    \label{tab:swot:tq-fm}}
  \footnotesize
  \begin{tabularx}{\textwidth}{XX}
    \toprule
    \textbf{Strengths} (Sec.~\ref{sec:strengths})
    \begin{enumerate}[S1.]
    \item Support for independent check of deductive proof objects 
    \item Introspective reasoning
    \item TQL-5/T2 qualification guidance for model checkers
    \item Safety \& liveness certificates for model checkers \& 
      solvers
    \item Verified model checking algorithms
    \item IEC~60880 qualification guidance for verified compilers
    \item TQL-1 qualification guidance for model transformers
    \item TQL-1 certification for DAL-A code generators
    \item T3 certification of diverse code generators
    \end{enumerate}
    &
    \textbf{Weaknesses} (Sec.~\ref{sec:weaknesses})
    \begin{enumerate}[W1.]
    \item Invalidation of theorems by user-defined axioms
    \item Mutual dependencies between checkers reduce diversity
    \item Undesired manual repetition of proof work
    \item Common errors in co-development of specs \& programs
    \item Common errors in tools and their artefact checkers
    \item Unqualified generators jeopardise artefact correctness 
    \item Standards incentivise incomplete testing of generators
    \item Established tools not yet considered intensively by industry
    \item Effort for validation of (translated) inputs
    \end{enumerate}
    \\\midrule
    \textbf{Opportunities} (Sec.~\ref{sec:opportunities})
    \begin{enumerate}[O1.]
    \item Motivation from enabling the use of proven components
    \item LCF approach with testable single soundness-critical core
    \item Guidance in manual proofs as fallback
    \item Provision of guidelines for addressing  specific standards
    \item Qualification of model checkers for TQL-5
    \item Semantics preserving compilers
    \item Code generation capabilities of proof assistants
    \item Proof-replay with diverse proof assistants
    \item Improve support for validation of (translated) inputs
    \item Expected increase in attention to qualification
    \end{enumerate}
    &
    \textbf{Threats} (Sec.~\ref{sec:threats})
    \begin{enumerate}[T1.]
    \item Size and complexity of code bases, even restricted to cores
    \item Critical components outside of the core
    \item Missing comprehensive documentation of service history
    \item Challenges in explaining indirect approaches
    \item Questionable necessity of lower levels if qualification 
    \item Size and complexity of proof objects for replay
    \item Undetected critical flaws in tool chains (e.g. compilers) 
    \item Lack of trust of certification assessors
    \item Discontinuities in tool maintenance
    \end{enumerate}
    \\\midrule
    \textbf{Qualification Requirements from the Standards} (Sec.~\ref{sec:tq-req})
    \begin{enumerate}[R1.]
    \item Tool configuration management
    \item Documentation of tool operational requirements 
    \item Verification of tool operational requirements 
    \item Validation of tool operational requirements
    \item Tool integration testing with an adequate test suite
    \item Comprehensive tool life cycle documentation
    \item Evidence of DAL-A-style verification activities
    \end{enumerate}
    &
      \textbf{Our Viewpoint and Suggestions} (Sec.~\ref{sec:what-needs-be})
    \begin{enumerate}[V1.]
    \item Put stronger focus on artefact-based tool qualification 
    \item Foster redundant proof results using diverse tools
    \item Further establish shared proof interchange formats
    \item Disclose qualification cases \& communicate them
    \item Improve/establish tool usage and proof auditing guidelines
    \item Further explore mutual proof object certification
    \item Aim at TQL-5 qualification of established model checkers
    \item Complement certified compilers with proof certificates
    \item Define qualification-preserving tool use cases
    \end{enumerate}
    \\\bottomrule
  \end{tabularx}
\end{table}

\section*{Acknowledgements}

Jan Peleska has been partially funded by the German Ministry of
Economics, Grant Agreement 20X1908E.
We would like to thank J\"org Brauer for helpful feedback and
highlighting the issue of tool suitability as well as Michael Leuschel
for his kind suggestions concerning EN~50128-based tool qualification.

\bibliography{main-arxiv}

\begin{thebibliography}{10}

\bibitem{Webster2014-GeneratingCertificationEvidence}
Matt Webster, Neil Cameron, Michael Fisher, and Mike Jump.
\newblock Generating certification evidence for autonomous unmanned aircraft
  using model checking and simulation.
\newblock {\em Journal of Aerospace Information Systems}, 11(5):258--279, 2014.

\bibitem{bedinfranca:inria-00551370}
Ricardo Bedin~Fran{\c c}a, Denis Favre-Felix, Xavier Leroy, Marc Pantel, and
  Jean Souyris.
\newblock {Towards Formally Verified Optimizing Compilation in Flight Control
  Software}.
\newblock In {\em Predictability and Performance in Embedded Systems},
  volume~18 of {\em OpenAccess Series in Informatics}, pages 59--68, Grenoble,
  France, March 2011. {Schloss Dagstuhl, Leibniz-Zentrum fuer Informatik}.

\bibitem{iec61508}
{IEC 61508}.
\newblock {\em Functional safety of electric/electronic/programmable electronic
  safety-related systems}.
\newblock {I}nternational {E}lectrotechnical {C}ommission, 2006.

\bibitem{DO178C}
RTCA SC-205/EUROCAE WG-71.
\newblock {\em {RTCA DO-178C -- Software Considerations in Airborne Systems and
  Equipment Certification}}.
\newblock 1140 Connecticut Avenue, N.W., Suite 1020, Washington, D.C. 20036,
  December 2011.

\bibitem{CENELEC50128}
{CENELEC}.
\newblock {\em {EN 50128:2011 Railway applications - Communication, signalling
  and processing systems - Software for railway control and protection
  systems}}.
\newblock 2011.

\bibitem{iso26262-8}
{ISO/DIS 26262-8}.
\newblock {\em Road vehicles -- functional safety -- Part 8: Supporting
  processes}, 2009.

\bibitem{DO330}
RTCA SC-205/EUROCAE WG-71.
\newblock {\em {RTCA DO-330 -- Software Tool Qualification Considerations}}.
\newblock 1140 Connecticut Avenue, N.W., Suite 1020, Washington, D.C. 20036,
  December 2011.

\bibitem{CENELEC50126a}
{CENELEC}.
\newblock {\em {EN 50126-1:2018 Railway Applications - The Specification and
  Demonstration of Reliability, Availability, Maintainability and Safety (RAMS)
  - Part 1: Generic RAMS Process}}.
\newblock 2018.

\bibitem{DBLP:journals/dagstuhl-reports/CoferKSW15}
Darren~D. Cofer, Gerwin Klein, Konrad Slind, and Virginie Wiels.
\newblock Qualification of formal methods tools (dagstuhl seminar 15182).
\newblock {\em Dagstuhl Reports}, 5(4):142--159, 2015.

\bibitem{Wagner2017-FormalMethodsTool}
Lucas~G. Wagner, Darren Cofer, Konrad Slind, Cesare Tinelli, and Alain Mebsout.
\newblock Formal methods tool qualification.
\newblock Technical Report NASA/CR-2017-219371, NASA, 2017.

\bibitem{DBLP:conf/cade/ChihaniMR13}
Zakaria Chihani, Dale Miller, and Fabien Renaud.
\newblock Foundational proof certificates in first-order logic.
\newblock In Maria~Paola Bonacina, editor, {\em Automated Deduction}, volume
  7898 of {\em LNCS}, pages 162--177. Springer, 2013.

\bibitem{Wagner2017-QualificationModelChecker}
Lucas Wagner, Alain Mebsout, Cesare Tinelli, Darren Cofer, and Konrad Slind.
\newblock Qualification of a model checker for avionics software verification.
\newblock In {\em NFM}, volume 10227 of {\em LNCS}, pages 404--419. Springer,
  Cham, 2017.

\bibitem{AET2013-SCADESuiteKCG}
{ANSYS Esterel Technologies}.
\newblock {SCADE} suite {KCG} 6.4 {DO-178C} certification kits technical data
  sheet.
\newblock Technical report, ANSYS, 2013.

\bibitem{kastner:hal-01399482}
Daniel K{\"a}stner, Xavier Leroy, Sandrine Blazy, Bernhard Schommer, Michael
  Schmidt, and Christian Ferdinand.
\newblock {Closing the Gap -- The Formally Verified Optimizing Compiler
  CompCert}.
\newblock In {\em {SSS'17: Safety-critical Systems Symposium 2017}},
  Developments in System Safety Engineering, pages 163--180, Bristol, United
  Kingdom, February 2017. {CreateSpace}.

\bibitem{DBLP:journals/jar/BlanchetteFLW18}
Jasmin~Christian Blanchette, Mathias Fleury, Peter Lammich, and Christoph
  Weidenbach.
\newblock A verified {SAT} solver framework with learn, forget, restart, and
  incrementality.
\newblock {\em J. Autom. Reason.}, 61(1-4):333--365, 2018.

\bibitem{Brauer2012-EfficientTrustworthyTool}
J{\"o}rg Brauer, Jan Peleska, and Uwe Schulze.
\newblock Efficient and trustworthy tool qualification for model-based testing
  tools.
\newblock In {\em Testing Software and Systems}, volume 7641 of {\em LNPSE},
  pages 8--23. Springer, Berlin Heidelberg, 2012.

\bibitem{Gleirscher2022-SoundDevelopmentSupervisors}
Mario Gleirscher, Lukas Plecher, and Jan Peleska.
\newblock Sound development of supervisors.
\newblock Working paper, U Bremen, 2022.

\bibitem{DBLP:journals/corr/abs-2003-06458}
Talia Ringer, Karl Palmskog, Ilya Sergey, Milos Gligoric, and Zachary Tatlock.
\newblock {QED} at large: {A} survey of engineering of formally verified
  software.
\newblock {\em CoRR}, abs/2003.06458, 2020.

\bibitem{Wimmer2019-MuntaVerifiedModel}
Simon Wimmer.
\newblock Munta: A verified model checker for timed automata.
\newblock In {\em FORMATS}, volume 11750 of {\em LNTCS}, pages 236--243.
  Springer, 2019.

\bibitem{DBLP:series/txtcs/BertotC04}
Yves Bertot and Pierre Cast{\'{e}}ran.
\newblock {\em Interactive Theorem Proving and Program Development - Coq'Art:
  The Calculus of Inductive Constructions}.
\newblock Texts in Theoretical Computer Science. An {EATCS} Series. Springer,
  2004.

\bibitem{DBLP:books/sp/NipkowPW02}
Tobias Nipkow, Lawrence~C. Paulson, and Markus Wenzel.
\newblock {\em Isabelle/HOL - {A} Proof Assistant for Higher-Order Logic},
  volume 2283 of {\em Lecture Notes in Computer Science}.
\newblock Springer, 2002.

\bibitem{Platzer2022-KeYmaeraXTutorial}
Andr{\'e} Platzer.
\newblock {KeYmaera X} tutorial.
\newblock Technical report, Computer Science Department, Carnegie Mellon
  University, 2022.

\bibitem{Moura2015-LeanTheoremProver}
Leonardo de~Moura, Soonho Kong, Jeremy Avigad, Floris van Doorn, and Jakob von
  Raumer.
\newblock The {Lean} theorem prover (system description).
\newblock In {\em Automated Deduction - {CADE}-25}, volume 9195 of {\em LNCS},
  pages 378--388. Springer, Cham, 2015.

\bibitem{Owre1992-PVSprototypeverification}
S.~Owre, J.~Rushby, and N.~Shankar.
\newblock {PVS}: A prototype verification system.
\newblock In {\em Automated Deduction (CADE-11)}, volume 607 of {\em LNCS},
  pages 748--752. Springer, 1992.

\bibitem{DBLP:journals/sttt/Gibson-Robinson16}
Thomas Gibson{-}Robinson, Philip~J. Armstrong, Alexandre Boulgakov, and A.~W.
  Roscoe.
\newblock {FDR3:} a parallel refinement checker for {CSP}.
\newblock {\em Int. J. Softw. Tools Technol. Transf.}, 18(2):149--167, 2016.

\bibitem{Champion2016-Kind2Model}
Adrien Champion, Alain Mebsout, Christoph Sticksel, and Cesare Tinelli.
\newblock The \textsc{Kind} 2 model checker.
\newblock In {\em Computer Aided Verification}, pages 510--517. Springer, 2016.

\bibitem{Cavada2014-nuXmvSymbolicModel}
Roberto Cavada, Alessandro Cimatti, Michele Dorigatti, Alberto Griggio,
  Alessandro Mariotti, Andrea Micheli, Sergio Mover, Marco Roveri, and Stefano
  Tonetta.
\newblock The {nuXmv} symbolic model checker.
\newblock In {\em Computer Aided Verification}, volume 8559 of {\em LNCS},
  pages 334--342. Springer, 2014.

\bibitem{Kwiatkowska2011-PRISM4.0Verification}
Marta Kwiatkowska, Gethin Norman, and David Parker.
\newblock {PRISM} 4.0: Verification of probabilistic real-time systems.
\newblock In G.~Gopalakrishnan and S.~Qadeer, editors, {\em Computer Aided
  Verification ({CAV}), 23rd Int. Conf.}, volume 6806 of {\em LNCS}, pages
  585--591. Springer, 2011.

\bibitem{Leuschel2003-ProBModelChecker}
Michael Leuschel and Michael Butler.
\newblock {ProB}: A model checker for b.
\newblock In {\em Formal Methods}, pages 855--874. Springer, Berlin Heidelberg,
  2003.

\bibitem{DBLP:books/daglib/0083128}
Pierre Deransart, AbdelAli Ed{-}Dbali, and Laurent Cervoni.
\newblock {\em Prolog - the standard: reference manual}.
\newblock Springer, 1996.

\bibitem{Holzmann1997-ModelCheckerSPIN}
Gerard~J. Holzmann.
\newblock The model checker {SPIN}.
\newblock {\em IEEE Trans. Software Eng.}, 23(5):279--295, 1997.

\bibitem{Behrmann2004-TutorialUppaal}
Gerd Behrmann, Alexandre David, and Kim~Guldstrand Larsen.
\newblock A tutorial on {\textsc{uppaal}}.
\newblock In {\em SFM}, volume 3185 of {\em LNCS}, pages 200--236, Berlin,
  Heidelberg, 2004. Springer.

\bibitem{Barnes2003-HighIntegritySoftware}
John Barnes.
\newblock {\em High Integrity Software}.
\newblock Addison Wesley, London, 2003.

\bibitem{SAEARP-47541996}
{SAE/ARP-4754}.
\newblock Certification considerations for highly-integrated or complex
  aircraft systems.
\newblock Standard, Society of Automotive Engineers (SAE) / Aerospace
  Recommended Practice (ARP), 1996.

\bibitem{Piercy1989-MakingSWOTAnalysis}
Nigel Piercy and William Giles.
\newblock Making {SWOT} analysis work.
\newblock {\em Marketing Intelligence \& Planning}, 7(5/6):5--7, 1989.

\bibitem{Garavel2020-2020ExpertSurvey}
Hubert Garavel, Maurice~H. ter Beek, and Jaco van~de Pol.
\newblock The 2020 expert survey on formal methods.
\newblock In {\em Formal Methods for Industrial Critical Systems}, pages 3--69.
  Springer, 2020.

\bibitem{Gleirscher2020-FormalMethodsDependable}
Mario Gleirscher and Diego Marmsoler.
\newblock Formal methods in dependable systems engineering: A survey of
  professionals from {Europe} and {North America}.
\newblock {\em Empirical Software Engineering}, 25(6):4473--4546, 2020.
\newblock Presented at the ESEC/FSE'21 journal first track.

\bibitem{DBLP:conf/tphol/BerghoferN00}
Stefan Berghofer and Tobias Nipkow.
\newblock Proof terms for simply typed higher order logic.
\newblock In Mark Aagaard and John Harrison, editors, {\em Theorem Proving in
  Higher Order Logics, 13th International Conference, TPHOLs 2000, Portland,
  Oregon, USA, August 14-18, 2000, Proceedings}, volume 1869 of {\em LNCS},
  pages 38--52. Springer, 2000.

\bibitem{wenzel2021isabelle}
Makarius Wenzel.
\newblock The {Isabelle/Isar} reference manual.
\newblock https://isabelle.in.tum.de/dist/doc/isar-ref.pdf, December 2021.

\bibitem{DBLP:journals/fmsd/Wong99}
Wai Wong.
\newblock Validation of {HOL} proofs by proof checking.
\newblock {\em Formal Methods Syst. Des.}, 14(2):193--212, 1999.

\bibitem{DBLP:conf/cade/NipkowR21}
Tobias Nipkow and Simon Ro{\ss}kopf.
\newblock Isabelle's metalogic: Formalization and proof checker.
\newblock In Andr{\'{e}} Platzer and Geoff Sutcliffe, editors, {\em Automated
  Deduction - {CADE} 28 - 28th International Conference on Automated Deduction,
  Virtual Event, July 12-15, 2021, Proceedings}, volume 12699 of {\em LNCS},
  pages 93--110. Springer, 2021.

\bibitem{DBLP:journals/pacmpl/SozeauBFTW20}
Matthieu Sozeau, Simon Boulier, Yannick Forster, Nicolas Tabareau, and
  Th{\'{e}}o Winterhalter.
\newblock {Coq} {Coq} correct! verification of type checking and erasure for
  {Coq}, in {Coq}.
\newblock {\em Proc. {ACM} Program. Lang.}, 4({POPL}):8:1--8:28, 2020.

\bibitem{DBLP:journals/jlap/Abrahamsson20}
Oskar Abrahamsson.
\newblock A verified proof checker for higher-order logic.
\newblock {\em J. Log. Algebraic Methods Program.}, 112:100530, 2020.

\bibitem{anssi_inria2021coq}
ANSSI-INRIA.
\newblock Requirements on the use of {Coq} in the context of {Common Criteria}
  evaluations.
\newblock {\em J. Log. Algebraic Methods Program.}, 2021.

\bibitem{DBLP:conf/cade/Harrison06}
John Harrison.
\newblock Towards self-verification of {HOL} light.
\newblock In Ulrich Furbach and Natarajan Shankar, editors, {\em Automated
  Reasoning, Third International Joint Conference, {IJCAR} 2006, Seattle, WA,
  USA, August 17-20, 2006, Proceedings}, volume 4130 of {\em LNCS}, pages
  177--191. Springer, 2006.

\bibitem{DBLP:journals/jar/KumarAMO16}
Ramana Kumar, Rob Arthan, Magnus~O. Myreen, and Scott Owens.
\newblock Self-formalisation of higher-order logic - semantics, soundness, and
  a verified implementation.
\newblock {\em J. Autom. Reason.}, 56(3):221--259, 2016.

\bibitem{DBLP:conf/itp/AbrahamssonMKS22}
Oskar Abrahamsson, Magnus~O. Myreen, Ramana Kumar, and Thomas Sewell.
\newblock Candle: {A} verified implementation of {HOL} light.
\newblock In June Andronick and Leonardo de~Moura, editors, {\em 13th
  International Conference on Interactive Theorem Proving, {ITP} 2022, August
  7-10, 2022, Haifa, Israel}, volume 237 of {\em LIPIcs}, pages 3:1--3:17.
  Schloss Dagstuhl - Leibniz-Zentrum f{\"{u}}r Informatik, 2022.

\bibitem{DBLP:conf/itp/Myreen21}
Magnus~O. Myreen.
\newblock The {CakeML} project's quest for ever stronger correctness theorems.
\newblock In Liron Cohen and Cezary Kaliszyk, editors, {\em Interactive Theorem
  Proving ({ITP}), 12th Int Conf}, volume 193 of {\em LIPIcs}, pages 1:1--1:10.
  Schloss Dagstuhl - Leibniz-Zentrum f{\"{u}}r Informatik, 2021.

\bibitem{Bendisposto2014-Whowatcheswatchers}
Jens Bendisposto, Sebastian Krings, and Michael Leuschel.
\newblock Who watches the watchers: Validating the {ProB} validation tool.
\newblock In C.~Dubois, D.~Giannakopoulou, and D.~Mry, editors, {\em F-IDE},
  volume 149 of {\em EPTCS}, pages 16--29, 2014.

\bibitem{ClearSy2009-AtelierBUser}
{ClearSy}.
\newblock {\em {Atelier B}: User and Reference Manuals}.
\newblock Aix-en-Provence, France, 2009.

\bibitem{Stump2012-SMTproofchecking}
Aaron Stump, Duckki Oe, Andrew Reynolds, Liana Hadarean, and Cesare Tinelli.
\newblock {SMT} proof checking using a logical framework.
\newblock {\em Form Method Syst Des}, 42(1):91--118, 2012.

\bibitem{Barrett2011-CVC4}
Clark Barrett, Christopher~L. Conway, Morgan Deters, Liana Hadarean, Dejan
  Jovanovi{\'{c}}, Tim King, Andrew Reynolds, and Cesare Tinelli.
\newblock {CVC}4.
\newblock In {\em CAV}, pages 171--177. Springer, 2011.

\bibitem{Mebsout2016-ProofcertificatesSMT}
Alain Mebsout and Cesare Tinelli.
\newblock Proof certificates for {SMT}-based model checkers for infinite-state
  systems.
\newblock In {\em Formal Methods in Computer-Aided Design ({FMCAD})}, pages
  117--124. {IEEE}, 2016.

\bibitem{Wimmer2020-VerifiedCertificationReachability}
Simon Wimmer and Joshua von Mutius.
\newblock Verified certification of reachability checking for timed automata.
\newblock In {\em TACAS}, volume 12078 of {\em LNCS}, pages 425--443. Springer,
  2020.

\bibitem{Griggio2021-CertifyingproofsSAT}
Alberto Griggio, Marco Roveri, and Stefano Tonetta.
\newblock Certifying proofs for {SAT}-based model checking.
\newblock {\em FMSD}, 57(2):178--210, 2021.

\bibitem{Wimmer2020-CertifyingEmptinessTimed}
Simon Wimmer, Fr{\'{e}}d{\'{e}}ric Herbreteau, and Jaco van~de Pol.
\newblock Certifying emptiness of timed {B\"uchi} automata.
\newblock In {\em FORMATS}, volume 12288 of {\em LNCS}, pages 58--75. Springer,
  2020.

\bibitem{Hoelzl2012-VerifyingpCTLModel}
Johannes H{\"o}lzl and Tobias Nipkow.
\newblock Verifying {pCTL} model checking.
\newblock In Cormac Flanagan and Barbara K{\"o}nig, editors, {\em TACAS},
  volume 7214 of {\em LNTCS}, pages 347--361, Berlin, Heidelberg, 2012.
  Springer.

\bibitem{Wimmer2018-VerifiedModelChecking}
Simon Wimmer and Peter Lammich.
\newblock Verified model checking of timed automata.
\newblock In {\em TACAS}, volume 10805 of {\em LNTCS}, pages 61--78. Springer,
  2018.

\bibitem{Schimpf2009-ConstructionBuchiAutomata}
Alexander Schimpf, Stephan Merz, and Jan-Georg Smaus.
\newblock Construction of {B{\"u}chi} automata for {LTL} model checking
  verified in isabelle/hol.
\newblock In Stefan Berghofer, Tobias Nipkow, Christian Urban, and Makarius
  Wenzel, editors, {\em TPHOL}, volume 5674 of {\em LNTCS}, pages 424--439,
  Berlin, Heidelberg, 2009. Springer.

\bibitem{kastner:hal-01643290}
Daniel K{\"a}stner, J{\"o}rg Barrho, Ulrich W{\"u}nsche, Marc Schlickling,
  Bernhard Schommer, Michael Schmidt, Christian Ferdinand, Xavier Leroy, and
  Sandrine Blazy.
\newblock {CompCert: Practical Experience on Integrating and Qualifying a
  Formally Verified Optimizing Compiler}.
\newblock In {\em {ERTS2 2018 - 9th European Congress Embedded Real-Time
  Software and Systems}}, pages 1--9, Toulouse, France, January 2018. {3AF,
  SEE, SIE}.

\bibitem{Taft2017-BuildingTrustModel}
S.~Tucker Taft, Elie Richa, and Andres Toom.
\newblock Building trust in a model-based automatic code generator.
\newblock {\em {ACM} {SIGAda} Ada Letters}, 36(2):54--57, 2017.

\bibitem{DBLP:conf/oopsla/Colaco20}
Jean{-}Louis Cola{\c{c}}o.
\newblock An overview of scade, a synchronous language for safety-critical
  software (keynote).
\newblock In {\em {REBLS} 2020: 7th {ACM} {SIGPLAN} International Workshop on
  Reactive and Event-Based Languages and Systems, Virtual Event, USA, November
  16, 2020}, page~1. {ACM}, 2020.

\bibitem{Leuschel2022-PrivateCommunication}
Michael Leuschel.
\newblock Private communication, September 2022.

\bibitem{Yuan2022-EndEndMechanized}
Shenghao Yuan, Fr{\'{e}}d{\'{e}}ric Besson, Jean-Pierre Talpin, Samuel Hym,
  Koen Zandberg, and Emmanuel Baccelli.
\newblock End-to-end mechanized proof of~an~{eBPF} virtual machine
  for~micro-controllers.
\newblock In {\em Computer Aided Verification}, pages 293--316. Springer, 2022.

\bibitem{Necula1998-designimplementationcertifying}
George~C. Necula and Peter Lee.
\newblock The design and implementation of a certifying compiler.
\newblock {\em {ACM} {SIGPLAN} Notices}, 33(5):333--344, 1998.

\bibitem{DBLP:journals/jfrea/Adams16}
Mark~Miles Adams.
\newblock Proof auditing formalised mathematics.
\newblock {\em J. Formaliz. Reason.}, 9(1):3--32, 2016.

\bibitem{DBLP:conf/esop/MonniauxB22}
David Monniaux and Sylvain Boulm{\'{e}}.
\newblock The trusted computing base of the {CompCert} verified compiler.
\newblock In Ilya Sergey, editor, {\em Programming Languages and Systems - 31st
  European Symposium on Programming ({ESOP}), held as part of {ETAPS}}, volume
  13240 of {\em LNCS}, pages 204--233. Springer, 2022.

\bibitem{DBLP:conf/vstte/Lawford16}
Mark Lawford.
\newblock Stupid tool tricks for smart model based design.
\newblock In Sandrine Blazy and Marsha Chechik, editors, {\em Verified
  Software. Theories, Tools, and Experiments - 8th International Conference,
  {VSTTE} 2016, Toronto, ON, Canada, July 17-18, 2016, Revised Selected
  Papers}, volume 9971 of {\em LNCS}, pages 1--7, 2016.

\bibitem{belanger2019certified}
Olivier~Savary B{\'e}langer, Matthew~Z Weaver, and Andrew~W Appel.
\newblock Certified code generation from {CPS} to {C}.
\newblock {\em Proc. ACM}, 2019.

\bibitem{DBLP:conf/esop/HupelN18}
Lars Hupel and Tobias Nipkow.
\newblock A verified compiler from {Isabelle/HOL} to {CakeML}.
\newblock In Amal Ahmed, editor, {\em Programming Languages and Systems},
  volume 10801 of {\em LNCS}, pages 999--1026. Springer, 2018.

\bibitem{Cofer2015-Youkeepusing}
Darren Cofer.
\newblock You keep using that word.
\newblock {\em {ACM} {SIGLOG} News}, 2(4):17--25, October 2015.

\bibitem{DBLP:conf/sosp/KleinEHACDEEKNSTW09}
Gerwin Klein, Kevin Elphinstone, Gernot Heiser, June Andronick, David Cock,
  Philip Derrin, Dhammika Elkaduwe, Kai Engelhardt, Rafal Kolanski, Michael
  Norrish, Thomas Sewell, Harvey Tuch, and Simon Winwood.
\newblock sel4: formal verification of an {OS} kernel.
\newblock In Jeanna~Neefe Matthews and Thomas~E. Anderson, editors, {\em
  Proceedings of the 22nd {ACM} Symposium on Operating Systems Principles
  2009}, pages 207--220. {ACM}, 2009.

\bibitem{Klein2018-Formallyverifiedsoftware}
Gerwin Klein, June Andronick, Matthew Fernandez, Ihor Kuz, Toby Murray, and
  Gernot Heiser.
\newblock Formally verified software in the real world.
\newblock {\em Commun. ACM}, 61(10):68--77, 2018.

\bibitem{VanderLeest2018}
Steven~H. VanderLeest.
\newblock Is formal proof of {seL4} sufficient for avionics security?
\newblock {\em IEEE Aerospace and Electronic Systems Magazine}, 33(2):16--21,
  2018.

\bibitem{bertot2020coq}
Yves Bertot, Maxime D{\'e}n{\'e}s, Vincent Laporte, Arnaud Fontaine, and Thomas
  Letan.
\newblock The use of {Coq} for {Common Criteria} evaluations.
\newblock In {\em {CoqPL 2020 The Sixth International Workshop on Coq for
  Programming Languages}}, pages 1--3, New Orleans, United States, 2020.

\bibitem{brahmi:hal-01708332}
Abderrahmane Brahmi, David Delmas, Mohamed~Habib Essoussi, Famantanantsoa
  Randimbivololona, Abdellatif Atki, and Thomas Marie.
\newblock {Formalise to automate: deployment of a safe and cost-efficient
  process for avionics software}.
\newblock In {\em {9th European Congress on Embedded Real Time Software and
  Systems (ERTS 2018)}}, pages 1--11, Toulouse, France, January 2018.

\bibitem{DBLP:conf/cav/EsparzaLNNSS13}
Javier Esparza, Peter Lammich, Ren{\'{e}} Neumann, Tobias Nipkow, Alexander
  Schimpf, and Jan{-}Georg Smaus.
\newblock A fully verified executable {LTL} model checker.
\newblock In Natasha Sharygina and Helmut Veith, editors, {\em Computer Aided
  Verification}, volume 8044 of {\em LNCS}, pages 463--478. Springer, 2013.

\bibitem{DBLP:conf/nfm/Hurd11}
Joe Hurd.
\newblock The opentheory standard theory library.
\newblock In Mihaela~Gheorghiu Bobaru, Klaus Havelund, Gerard~J. Holzmann, and
  Rajeev Joshi, editors, {\em {NASA} Formal Methods - Third International
  Symposium, {NFM} 2011, Pasadena, CA, USA, April 18-20, 2011. Proceedings},
  volume 6617 of {\em LNCS}, pages 177--191. Springer, 2011.

\bibitem{DBLP:journals/jsc/GauthierK19}
Thibault Gauthier and Cezary Kaliszyk.
\newblock Aligning concepts across proof assistant libraries.
\newblock {\em J. Symb. Comput.}, 90:89--123, 2019.

\bibitem{DBLP:conf/itp/KellerW10}
Chantal Keller and Benjamin Werner.
\newblock Importing {HOL} light into {Coq}.
\newblock In Matt Kaufmann and Lawrence~C. Paulson, editors, {\em Interactive
  Theorem Proving, First International Conference, {ITP} 2010, Edinburgh, UK,
  July 11-14, 2010. Proceedings}, volume 6172 of {\em LNCS}, pages 307--322.
  Springer, 2010.

\bibitem{Farrell2022-FRETtingRequirementsFormalised}
Marie Farrell, Matt Luckcuck, Ois{\'{\i}}n Sheridan, and Rosemary Monahan.
\newblock {FRETting} about requirements: Formalised requirements
  for~an~aircraft engine controller.
\newblock In {\em REFSQ}, pages 96--111. Springer, 2022.

\bibitem{Foster2020-IntegrationFormalProof}
Simon Foster, Yakoub Nemouchi, Mario Gleirscher, Ran Wei, and Tim Kelly.
\newblock Integration of formal proof into unified assurance cases with
  {Isabelle/SACM}.
\newblock {\em Form Asp Comput}, 2021.

\bibitem{Chockler2006-CoverageMetricsFormal}
Hana Chockler, Orna Kupferman, and Moshe Vardi.
\newblock Coverage metrics for formal verification.
\newblock {\em Int. J. Softw. Tools Technol. Trans.}, 8(4):373--386, 2006.

\bibitem{Brucker2019-IsabelleDOFDesignImplementation}
Achim~D. Brucker and Burkhart Wolff.
\newblock Isabelle/{DOF}: Design and implementation.
\newblock In {\em SEFM}, volume 11724 of {\em LNTCS}, pages 275--292. Springer,
  2019.

\bibitem{DBLP:journals/entcs/Wiedijk12}
Freek Wiedijk.
\newblock Pollack-inconsistency.
\newblock {\em Electron. Notes Theor. Comput. Sci.}, 285:85--100, 2012.

\bibitem{DBLP:conf/itp/Adams16}
Mark Adams.
\newblock {HOL} zero's solutions for pollack-inconsistency.
\newblock In Jasmin~Christian Blanchette and Stephan Merz, editors, {\em
  Interactive Theorem Proving - 7th International Conference, {ITP} 2016,
  Nancy, France, August 22-25, 2016, Proceedings}, volume 9807 of {\em LNCS},
  pages 20--35. Springer, 2016.

\bibitem{Yang2011-Findingunderstandingbugs}
Xuejun Yang, Yang Chen, Eric Eide, and John Regehr.
\newblock Finding and understanding bugs in {C} compilers.
\newblock In {\em Programming language design and implementation ({PLDI}), 32nd
  {ACM} {SIGPLAN} Conf}, pages 283--294. {ACM} Press, 2011.

\bibitem{Barendregt2005-challengecomputermathematics}
Henk Barendregt and Freek Wiedijk.
\newblock The challenge of computer mathematics.
\newblock {\em Philosophical Transactions of the Royal Society A: Mathematical,
  Physical and Engineering Sciences}, 363(1835):2351--2375, September 2005.

\bibitem{DBLP:journals/sttt/PnueliSS98}
Amir Pnueli, Ofer Strichman, and Michael Siegel.
\newblock The code validation tool {CVT:} automatic verification of a
  compilation process.
\newblock {\em Int. J. Softw. Tools Technol. Transf.}, 2(2):192--201, 1998.

\bibitem{DBLP:conf/birthday/PnueliSS99}
Amir Pnueli, Ofer Strichman, and Michael Siegel.
\newblock Translation validation: From {SIGNAL} to {C}.
\newblock In Ernst{-}R{\"{u}}diger Olderog and Bernhard Steffen, editors, {\em
  Correct System Design, Recent Insight and Advances, (to Hans Langmaack on the
  occasion of his retirement from his professorship at the University of
  Kiel)}, volume 1710 of {\em Lecture Notes in Computer Science}, pages
  231--255. Springer, 1999.

\bibitem{Gleirscher2021-ManifestoApplicableFormal}
Mario Gleirscher, Jaco van~de Pol, and James Woodcock.
\newblock A manifesto for applicable formal methods.
\newblock Working paper, University of Bremen and Aarhus University and
  University of York, 2021.

\bibitem{Fantechi2011-AdoptionModelChecking}
Alessandro Fantechi and Stefania Gnesi.
\newblock On the adoption of model checking in safety-related software
  industry.
\newblock In {\em SAFECOMP}, volume 6894 of {\em LNPSE}, pages 383--396.
  Springer, Berlin Heidelberg, 2011.

\bibitem{DBLP:conf/rssrail/ParillaudFB19}
Camille Parillaud, Yoann Fonteneau, and Fabien Belmonte.
\newblock Interlocking formal verification at alstom signalling.
\newblock In Simon~Collart Dutilleul, Thierry Lecomte, and Alexander~B.
  Romanovsky, editors, {\em Reliability, Safety, and Security of Railway
  Systems}, volume 11495 of {\em Lecture Notes in Computer Science}, pages
  215--225. Springer, 2019.

\bibitem{DBLP:journals/jar/Sutcliffe17}
Geoff Sutcliffe.
\newblock The {TPTP} problem library and associated infrastructure - from {CNF}
  to th0, {TPTP} v6.4.0.
\newblock {\em J. Autom. Reason.}, 59(4):483--502, 2017.

\bibitem{Beyer2016-CorrectnessWitnessesExchanging}
Dirk Beyer, Matthias Dangl, Daniel Dietsch, and Matthias Heizmann.
\newblock Correctness witnesses: Exchanging verification results between
  verifiers.
\newblock In {\em Foundations of Software Engineering (FSE), 24th ACM SIGSOFT
  Int Symp}, pages 326--337. ACM, 2016.

\bibitem{Ibrahim2021-StateArtSoftware}
Mohamad Ibrahim and Umut Durak.
\newblock State of the art in software tool qualification with {DO}-330: A
  survey.
\newblock In S.~G\"otz, L.~Linsbauer, I.~Schaefer, and A.~Wortmann, editors,
  {\em Software Engineering ({SE}) Satellite Events}, volume 2814 of {\em LNI,
  CEUR Workshop Proceedings}, pages 1--22. 2021.

\end{thebibliography}
 \end{document}